# The Cosmic Web of Baryons

A White Paper submitted to *Galaxies across Cosmic Time (GCT)* and *The Cosmology and Fundamental Physics (CFP)* Science Frontiers Panels

Joel N. Bregman
Department of Astronomy
University of Michigan
Ann Arbor, MI    48109-1042
Email:   jbregman@umich.edu
Telephone:   734-764-3454

Claude R. Canizares, Massachusettes Institute of Technology
Reynue Cen: Princeton University
Jan-Willem den Herder:    SRON Netherlands Institute for Space Research
Massimiliano Bonamente:    University of Alabama in Huntsville
Taotao Fang:    University of California, Irvine
Massimiliano Galeazzi: University of Miami
Edward Jenkins:    Princeton University
Jelle S. Kaastra:    SRON Netherlands Institute for Space Research
Fabrizio Nicastro:    Harvard Smithsonian Center for Astrophysics
Smita Mathur:    Ohio State University
Takaya Ohashi:    Tokyo Metropolitan University
Frtis Paerels:    Columbia University
Kenneth Sembach:    Space Telescope Science Institute
Norbert Schulz:    Massachusettes Institute of Technology
Blair Savage:    University of Wisconsin
Randall Smith:    Harvard Smithsonian Center for Astrophysics
Noriko Yamasaki:    ISAS/JAXA
Bart Wakker:    University of Wisconsin

*1. The Science Goals*

Ordinary matter (baryons) represents ≈4.6% of the total mass/energy density of the Universe but less than 10% of this matter appears in collapsed objects (stars, galaxies, groups; Fukugita & Peebles 2004). Theory predicts that most of the baryons reside in vast unvirialized filaments that connect galaxy groups and clusters (the "Cosmic Web"; Fig 1b). After reionization, the dominant heating mechanism is through the shocks that develop when large-scale density waves collapse in the dark matter. As large-scale structure becomes more pronounced with cosmological time, the gas is increasingly shock-heated, reaching temperatures of $10^{5.5}$-$10^7$ K for z < 1 (Fig. 1a). Additional heating occurs through star-formation driven galactic winds and AGN, processes that pollute the surroundings with metals.

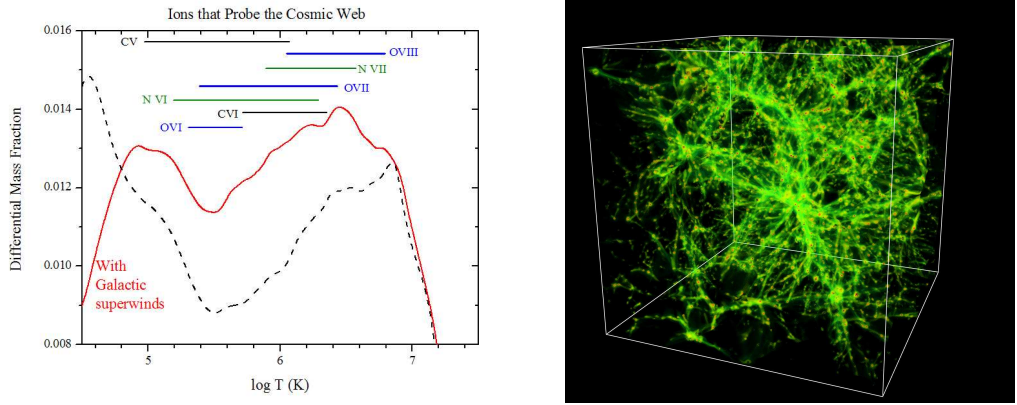

**Figure 1.** (left panel) The differential gas mass fraction at as a function of temperature at low redshift for the ΛCDM cosmological simulation of Cen & Ostriker (2006). This distribution is sensitive to the presence of galactic superwinds (solid red line; dashed line is without superwinds). The ions with the strongest resonance lines in the $10^5$-$10^7$ K range are shown, and except for OVI (UV line; 1035 Å), the other lines lie in the X-ray band. (right panel) The density distribution of baryons at low redshift from the same simulations. Most of the mass of the WHIM lies within the filaments that connect the higher density regions.

Ly α studies and OVI absorption line studies detect warm baryons, but ~50% of the baryons remain unaccounted for (Danforth & Shull 2008). These "missing baryons" can only be observed through X-ray studies. Therefore, a basic goal is to

- ***Determine if the missing baryons exist in the predicted hot phase***.

In the standard cosmology, chemical enrichment of the IGM occurs through galactic superwinds, a powerful feedback mechanism that also heats the gas. The shocks and superwinds leave distinctive features on absorption lines, such as double-lines (for a line of sight passing through a galactic superwind shell) and turbulent broadening. This feedback mechanism not only extends the cross section of the metal-enhanced regions, its effects are ion-dependent. An example of this is shown below (Figure 2; from Cen and Fang 2005), where there are significant differences between the three oxygen ions.

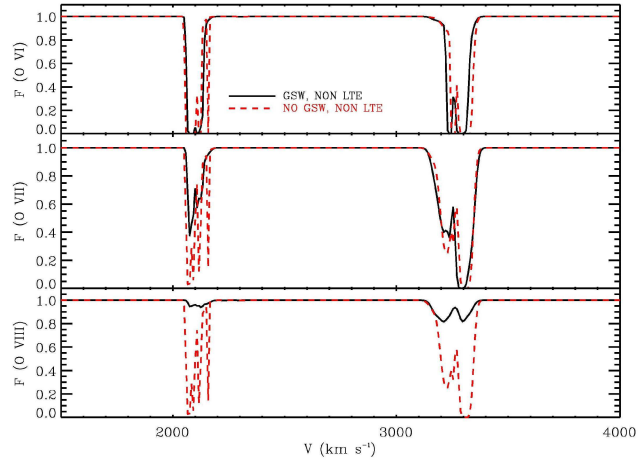

**Figure 2.** The absorption resulting from a line of sight through the simulation of Cen and Fang (2005; above), with galactic superiwnds (black) and without (red). The superwind region creates a double line absorption profile in this case and the OVIII lines are greatly enhanced.

These observational diagnostics allow us to pursue our second science goal:

- ***Test the large-scale structure and galactic superwind heating of the Cosmic Web***

The extent of the superwinds and the elemental mixing can be determined by studying the spatial relationship between hot gas seen through X-ray absorption and the location of galaxies (Stocke et al. 2006). By studying the same sight lines with X-rays and UV-optical bands, we will discover the relationship of all temperature components to the galactic environment. Therefore, X-ray studies will

- ***Measure the extent of galactic superwinds and the chemical mixing process.***

Finally, the baryons lie in Cosmic Web filaments extending between groups and clusters at 1-2 $R_{virial}$, as seen in Figure 1. Current surface brightness measurements rarely extend beyond $0.7R_{virial}$, yet they already suggest examples of the expected phenomenon (e.g., Werner et al. 2008). Studies of these structures become feasible when the limiting X-ray surface brightness can be lowered by an order of magnitude. This improvement is now possible, so our last goal is achievable:

- ***Measure the connections of Cosmic Web filaments to galaxy groups and clusters.***

*2. Observations Needed to Achieve these Goals*

The first three goals are achievable by measuring the He-like and H-like X-ray resonance lines of carbon (C V, C VI), nitrogen (N VI, N VII) and oxygen (O VII, OV III) toward background AGNs (other ions may be detected in a few cases, such as Ne IX, Ne X, Fe XVII, and Fe XVIII). Existing measurements of intergalactic OVII and OVIII are near current instrumental detection thresholds and therefore need confirmation (Nicastro et al. 2005; Kaastra et al. 2006; Bregman 2007; Buote et al.

2009) but the adjacent ion, intergalactic OVI, is detected in the UV along many sightlines and there are clear detections of OVII and OVIII within the Local Group. A conservative estimate of the equivalent width distribution, dN/dz, is obtained from models normalized to the OVI measurements (Cen and Fang 2006, Figure 3; the quality of the OVI normalization will improve significantly with upcoming COS observations). These show that we need an order of magnitude improvement over current sensitivities to conduct an X-ray survey of intergalactic absorption lines from the above elements. In addition, to study the velocity structures of lines, we need resolution that approaches the Doppler width of a line, typically 50-100 km/s, since superwinds from galaxies occur near the escape velocity, 200-1000 km/s.

There is great value to measuring multiple ionic states for understanding the temperature and relative abundance of the gas. With two ionic states, one can infer a temperature, provided that the line structures are the same. With three ions, one can begin to constrain a multi-temperature medium, should one exist (and differences in line structures will be an asset such determinations). From the column densities of three oxygen ions, OVI, OVII, and OVIII, we can test the predicted baryon temperature distributions (Figure 1) in a relative sense, without the need for absolute abundances. Finally, from relative abundance ratios of the elements, we can identify the types of supernovae that has heated and polluted the gas.

Absolute abundance determinations are challenging but possible in some cases. Some OVI abundances have been estimated from UV observations, and for OVII and OVIII measurements along the same lines of sight, their abundances can be estimated as well. Another approach is to obtain absorption line measurements through regions that have emission line detections, around clusters or groups (0.7-2 $R_{virial}$). The emission measure contains the electron density while the absorption is proportional to the ionic column, so by combining them, one can determine abundances.

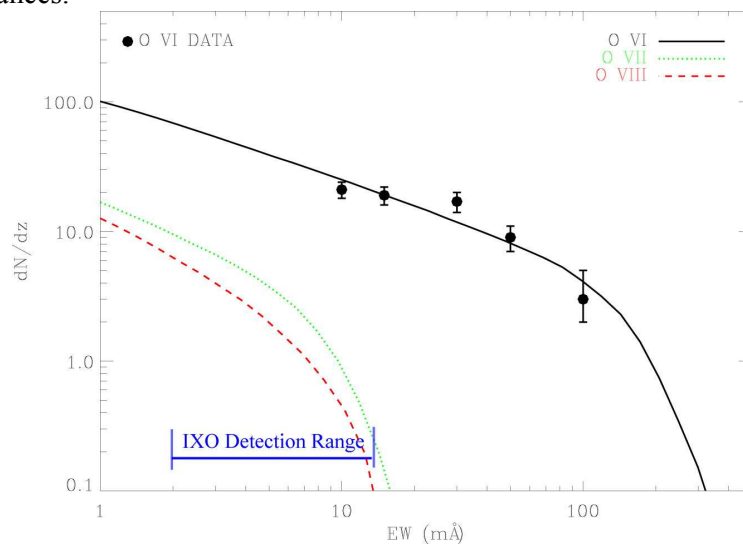

**Figure 3.** The differential number of absorbers as a function of equivalent width for OVI (λ1035), OVII (Kα), and OVIII (Kα), based on the model of Cen & Fang (2006); the OVI data are from Danforth and Shull (2005). Current claims of OVII and OVIII absorption are well above predictions and would indicate a temperature distribution significantly different than these models. The OVII and OVIII absorption lines predicted from the models, such as those in the range 2-15 mÅ, are not accessible to *Chandra* or *XMM*, but can be measured with *IXO*.

The primary observational program will obtain spectra for 30 known background AGNs, with a median flux of $5\times10^{-11}$ erg cm$^{-2}$ s$^{-1}$ and median redshift of

0.3. Against these AGNs, high-throughput spectroscopy in the 0.3–1.0 keV band will measure the abundant ion stages of heavy elements in the hot IGM (Figure 4). This program will address the hot baryons primarily in the low-redshift universe (z < 0.5), which is where the hot gas is predicted to be most common. Also, at low redshift, one can identify and study the galaxies at every detection site.

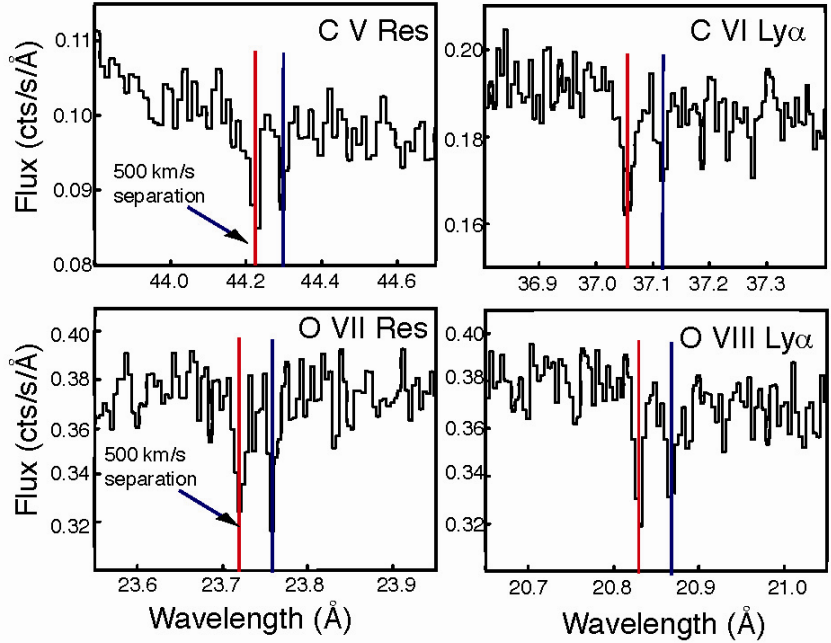

**Figure 4.** Simulations for *IXO*, with the baseline configuration, of the absorption by the Cosmic Web seen against a background AGN with a 0.5-2 keV flux of $5\times10^{-11}$ erg cm$^{-2}$ s$^{-1}$ and an exposure time of 600 ksec ($z_{abs}$ = 0.1); this is the 16$^{th}$ brightest AGN in a sample of 30. There is a multi-temperature gas (as given by Figure 1) with a superwind, producing a double-lined configuration with a separation of 500 km/s. For an AGN at an emission redshift of 0.3, calculations predict one absorption system with the strength shown. The lines shown here would be the signature of a multi-temperature medium with a temperature range of nearly an order of magnitude; additional lines are also detected (NVI, NVII).

The *International X-ray Observatory* (*IXO*) will have a sensitivity to the OVII resonance line that is 15 times better than with XMM or Chandra. The grating spectrometer has a baseline design resolution of 3000 (100 km/s) and a collecting area of 1000 cm$^2$. With *IXO*, the brightest quartile of sources can be studied in 2 Msec and the entire sample will require 18 Msec, a multi-year project (if the project achieves its goal of 3000 cm$^2$ in collecting area for the gratings, these times are reduced by a factor of three). The data products from this work will be absorption line measurements for the He-like and H-like species of O, C, and N, which exist in the temperature range $10^5$-$10^7$ K. Simulations show that the OVII line will be the strongest, but that multiple lines will be detected in about 80% of the targets. For each line, we obtain the column density (most lines are optically thin), velocity, and velocity width. We project that we will measure at least 100 OVII lines down to a limit of $10^{14.5}$ cm$^{-2}$; this will be sufficient to achieve our scientific goals. Potentially, this project can be extended to lower flux levels (especially with a collecting area of 3000 cm$^2$) since there are 146 sources in the *ROSAT* Bright Source Catalog with $|b|$ > 15° and $F_x$ > $2\times10^{-11}$ erg cm$^{-2}$ s$^{-1}$, most of which are AGNs. With 100 or more absorption line systems at low redshifts, it will be possible to construct correlation functions to compare to cosmological models. This has been carried out with optical-UV quasar absorption line systems and this study would be complementary in that it samples a hotter medium with a greater predicted overdensity.

In addition to the grating spectrometer, there is a quantum microcalorimeter planned for *IXO*, with much more collecting area but far lower spectral resolution (30,000 cm$^2$, resolution of 250 for the typical absorption line system). Our simulations show that it will be difficult to detect most weak lines with the microcalorimeter (Figure 5). The lines would be unresolved with the microcalorimeter and small gain variations in the instrument could masquerade as lines. However, the microcalorimeter is essential for detecting faint emission, as described below.

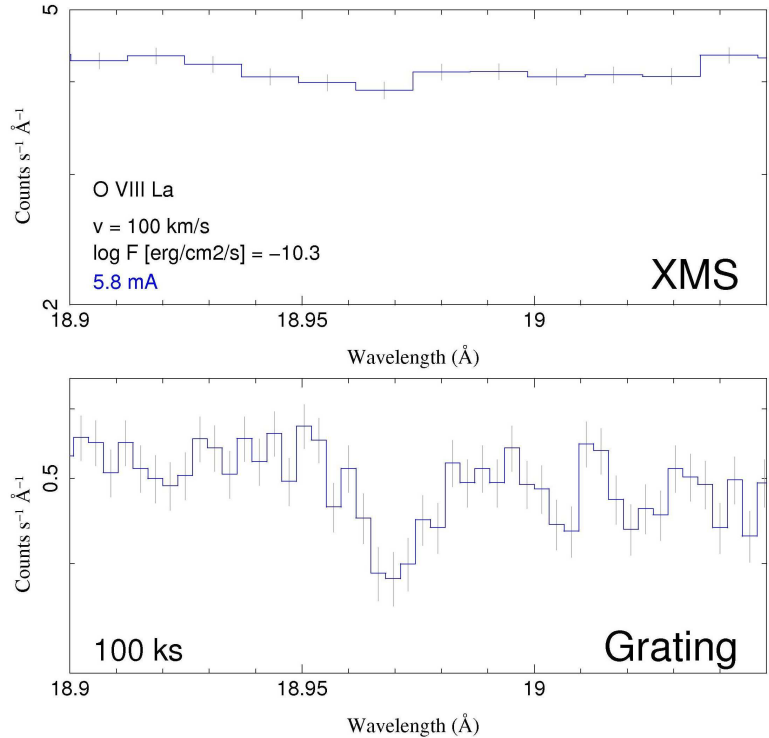

**Figure 5**. A simulation comparing the grating instrument to the microcalorimeter (XMS) shows the line clearly detected with the grating. Although the same line is formally detected in the XMS, it might be confused with small gain variations in the instrument.

The last goal requires imaging of Cosmic Web filaments as they extend away from galaxy groups and clusters. Such emission has been detected in a few unusual cases, such as the filament connecting Abell 222 and Abell 223 (Werner et al. 2008). However, in these cases, the overdensity (150-300) is somewhat higher than the typical predicted Cosmic Web filaments (10-200 for the hot gas).

From models, the temperature of gaseous filaments extending from clusters is typically $0.2$-$1 \times 10^7$ K, depending on the location in the filament (and on the model), so the bandpass used will be 0.3-1.5 keV. These observations are background-limited, and at an energy of about 0.5 keV, much of the background is due to line emission within the Milky Way. An enormous improvement in sensitivity can be realized by observing between the strong Milky Way lines. Such spectral imaging observations are possible with the quantum microcalorimeter, where the strong Milky Way emission lines are resolved and can be excluded. Also, *IXO* has sufficient angular resolution to remove most point sources (which do not dominate the background at 0.5 keV) and it has greater collecting area than previous instruments (by a factor of 15 at the relevant energies). The target clusters and groups must have a redshift great enough that the emission lines from the filament are shifted away from the strong Galactic emission lines (e.g., OVII). The optimum redshift ranges are 0.04-0.12 and 0.18-0.28. At these redshifts, target clusters will require mosaic observations to cover the outer regions where the filaments emerge. Even with these advantages, we are unlikely to reach overdensities down to 30, which will be the province of *Gen-X*. *IXO* should be able to measure emission from gas with overdensities near or below 100, which is beyond the classical virial radius. In practice, filaments extending from the clusters will have densities higher than the mean overdensity at that radius (closer to 100), which will aid in detection. *IXO* can be used to search for such filaments in the regions from 0.7-2 $R_{virial}$ with an observing time of about 1 Msec per cluster. A sensible program would begin by observing three clusters and enlarge the program as warranted by the results.